%% file: main.tex
\documentclass[a4paper,11pt]{article}

\usepackage{mathptmx}
\usepackage{multicol}
\usepackage{geometry}
\usepackage[english]{babel}
\usepackage{amsmath}
\usepackage{textcomp,amssymb}
\usepackage{setspace}
\usepackage{enumerate}
\usepackage{algorithm2e}
\usepackage{algpseudocode}
\usepackage{graphics}
\usepackage{hyperref}
\usepackage{booktabs}
\usepackage{xparse} 
\usepackage{
  color,
  float,
  epsfig,
  wrapfig,
  graphics,
  graphicx,
  subcaption
}
\usepackage{
  tikz,
  pgfplots,
  pgfplotstable
}

\usepackage{xspace}
\usepackage{multirow}
\usepackage{latexsym,fancyhdr,url}
\usepackage{listings}
\usepackage{pifont}
\newcommand{\cmark}{\ding{51}}
\newcommand{\xmark}{\ding{55}}
\newcommand{\ie}{\emph{i.e.,}\xspace}

\newcommand{\sys}{\mbox{\textsc{ChatFuzz}}\xspace}

\title{Augmenting Greybox Fuzzing with Generative AI}
\date{}
\begin{multicols}{3}
\author{Jie Hu\\University of California Riverside\\Riverside, CA 92521, USA\\jhu066@ucr.edu
    \and 
        Qian Zhang\\University of California Riverside\\Riverside, CA 92521, USA\\qzhang@cs.ucr.edu
    \and 
        Heng Yin\\University of California Riverside\\Riverside, CA 92521, USA\\heng@cs.ucr.edu }
\end{multicols}

\begin{document}
\maketitle
\input{0_abs}
\input{1_intro}

\input{2_background}
\input{3_method}

\input{4_results}
\input{5_discuss}
\input{6_related}

\input{7_conclusion}

\bibliographystyle{plain}
\bibliography{bib}
\end{document}

%% file: 0_abs.tex
\begin{abstract}

Real-world programs expecting structured inputs often has a format-parsing stage gating the deeper program space. Neither a mutation-based approach nor a generative approach can provide a solution that is effective and scalable. Large language models (LLM) pre-trained with an enormous amount of natural language corpus have proved to be effective for understanding the implicit format syntax and generating format-conforming inputs. In this paper, propose \sys, a greybox fuzzer augmented by generative AI. More specifically, we pick a seed in the fuzzer's seed pool and prompt ChatGPT generative models to variations, which are more likely to be format-conforming and thus of high quality. We conduct extensive experiments to explore the best practice for harvesting the power of generative LLM models. The experiment results show that our approach improves the edge coverage by 12.77\% over the SOTA greybox fuzzer (AFL++) on 12 target programs from three well-tested benchmarks. As for vulnerability detection, \sys is able to perform similar to or better than AFL++ for programs with explicit syntax rules but not for programs with non-trivial syntax.

\end{abstract}

%% file: 1_intro.tex
\section{Introduction}
\label{sec:intro}

In recent years, fuzz testing has emerged as an effective technique for testing software systems. For example, fuzz testing has been remarkably successful in uncovering critical security bugs in applications such as Chrome web-browser~\cite{chromiumbugs} and SQLLite database~\cite{sqllightbugs}. Generally, fuzz testing runs a program with seed inputs, mutates the previous inputs to improve a given guidance metric such as branch coverage, and repeats {this cycle of input mutation and the target program execution}. 

During the fuzzing process, we often execute the target program with generated large amount of test cases and monitor the runtime behavior to find vulnerabilities. For that, it is essential to generate test cases that effectively cover a wide range of execution paths and program behaviors. This comprehensive coverage enables thorough exploration of the program's functionality and helps uncover potential vulnerabilities or issues. The simplicity of fuzzing has made it a de-facto testing procedure for large-scale software systems; however, its effectiveness is based on an inherent yet oversighted assumption: a set of arbitrary input mutations is likely to yield meaningful inputs. In fact, our extensive experience suggests that this assumption often does not hold for most software systems that take highly structured data as inputs. 

\smallskip
\vspace{0.5pt}\noindent\textit{\textbf{Problem.}} In order to explore program states along the execution path, it is imperative to ensure that test inputs conform to a valid format. This is crucial because any input with a broken format would be promptly rejected at the program's format-parsing stage, limiting the exploration of deeper program states. There have been significant research efforts to ease the generation of highly structured data. Existing work for realistic input generation can be categorized into two groups: mutation-based approach and generative approach. Even with the most state-of-the-art techniques, this task of generating structured inputs can still be challenging across both categories.

With a mutation-based approach, the program inputs are generated by mutating parts of an existing seed. Greybox fuzzers such as American Fuzzy Lop (AFL)~\cite{afl} and its variations~\cite{lyu2019mopt, bohme2017directed, bohme2016coverage, fioraldi2020afl++}, whitebox fuzzers such as SAGE~\cite{godefroid2012sage}, SymSan~\cite{chen2022symsan}, as well as hybrid fuzzers like QSYM~\cite{yun2018qsym} fall under this category. The problem with mutation-based approach is that they often generate a large number of syntactically invalid inputs. We inspect the seeds generated in 2h by AFL++~\cite{fioraldi2020afl++} and SymSan for an XML parser, libxml2\footnote{https://gitlab.gnome.org/GNOME/libxml2.git}. The ratio of valid XML documents is 0.82\% and 16.93\% respectively. That is to say, deeper program states beyond format parsing stage is likely under-explored.

Prior works utilizing the generative approach construct structured input from scratch based on the specified grammar of the prospective testcase~\cite{yang2011finding, jsfunfuzz, han2019codealchemist, godefroid2008grammar}. However, it takes manual effort to construct the grammar of each different data format separately, which is time-consuming and often leads to incomplete grammar. To address this issue, learning based approach such as TreeFuzz~\cite{patra2016learning} have been proposed. Notably, 96.3\% of the JavaScript programs generated by TreeFuzz conform to the language's syntax. However, the generative approach overlooks feedback from the program's runtime to improve the input generate process. The exploration of program states during testing may be incomplete, because such generative approaches do not leverage runtime information to refine the input generation.

\medskip\noindent\textit{\textbf{This Work.}} The use of pre-trained large language models (LLMs) for text generation has become increasingly popular. LLMs have also been effective at translating natural language specifications and instructions to concrete code. In this paper, we lay the groundwork for embodying an AI-empowered fuzz testing technique to generate valid and realistic inputs. We observe that LLMs are capable of understanding the structures of provided examples and generating variants while retaining the same structure. This observation forms the foundation of AI-based input mutations.

Based on this observation, we propose a new fuzzing framework that augments greybox fuzzing with generative AI. We further implement a prototype system called \sys. \sys works as follows: in addition to the typical workflow of greybox fuzzing, a chat mutator picks a seed from the seed pool and prompts a ChatGPT model endpoint for similar inputs. These inputs are then evaluated by the greybox fuzzer, and some of them will be retained as new seeds for further exploration. 

The quality of the seed generated from the chat mutator is heavily dependent on the configuration of the ChatGPT model. We identify five hyper-parameters correlated to the performance of the chat mutator: 1) model endpoint choice; 2) model prompt design; 3) model response length; 4) the number of completion choices; and 5) the sampling temperature. We conduct experiments to explore different hyper-parameter configurations seeking to find the best practice for the chat mutator. 

To evaluate the efficacy of our new approach, we implemented a prototype \sys atop AFL++ and ChatGPT. We evaluated \sys on 12 programs expecting text-based input. Evaluation result demonstrates \sys on average improve edge coverage by 12.77\% compared to the original AFL++. Result analysis shows that \sys on average contributes to 12.77\% of the seeds in fuzzer queue. And \sys performs particularly well for programs expecting explicit syntax rather than non-trivial syntax.

\medskip
\noindent{\bf Contributions.} In this paper, we make the following contributions:

\begin{itemize}
    \item We identify the generation of the format-conforming input to be one major limitation of both the mutation-based approach and the generative approach.
    \item We propose to incorporate generative AI into a greybox fuzzer and provide detailed experiments exploring the best practice.
    \item We evaluate \sys to prove the effectiveness of our new approach. 
    \item We will open-source \sys along with datasets at \url{https://www.github.com/xxxx} upon acceptance of this work.
\end{itemize}

%% file: 2_background.tex
\section{Background and Related Work}
\label{sec:relwork}

\subsection{Large Language Models} 

Pre-trained LLMs are one type of neural networks with a huge amount of parameters. Those models are typically on a large corpus of text data in an {\em autoregressive} manner\textemdash trained to predict the next token in a text sequence. Such extensive pre-training enables them to function as {\em one-shot} or {\em zero-shot} learners~\cite{brown2020language}. In other words, those models can perform a variety of tasks when given only one example of the task or natural language-based instructions. The natural language instructions, along with any additional input data, that are passed to the LLMs are called {\em prompts}. 

\subsection{A Motivating Example}


\begin{table*}[t]
\centering
\caption{ChatGPT Example for XML}
\label{tab:gptsay}
\begin{tabular}{p{0.25\linewidth}p{0.25\linewidth}p{0.4\linewidth}}

\hline
\multirow{2}{*}{\textbf{Sample Input}} & \multicolumn{2}{c}{\textbf{ChatGPT Response Examples}}         \\ \cline{2-3} 
                                       & Example 1    & \multicolumn{1}{c}{Example 2} \\ \hline

\vspace{-0.120in}
\begin{lstlisting}
<doc>
  <clean> </clean>
  <dirty> A B </dirty>
  <mixed>
     A
     <clean> </clean>
     B
     <dirty> A B </dirty>
     C
  </mixed>
</doc>
\end{lstlisting}\vspace{-0.40in} &
\vspace{-0.120in}
\begin{lstlisting}
<doc>
  <clean> </clean>
  <dirty> X Y Z </dirty>
  <mixed>
    X
    <clean> </clean>
    Y
    <dirty> X Y Z </dirty>
    Z
  </mixed>
</doc>
\end{lstlisting}\vspace{-0.40in} & 
\vspace{-0.120in}
\begin{lstlisting}
<?xml version="1.0" encoding="UTF-8"?>
<books>
  <book>
    <title>The Great Gatsby</title>
    <author>F. Scott Fitzgerald</author>
    <genre>Drama</genre>
  </book>
  <book>
    <title>Pride and Prejudice</title>
    <author>Jane Austen</author>
    <genre>Romance</genre>
  </book>
</books>
\end{lstlisting}\vspace{-0.40in} \\ \hline
\end{tabular} 
\vspace{-0.20in}
\end{table*}

Previous research has explored a number of LLMs trained extensively on massive data. ChatGPT~\cite{vaswani2017attention} is one of them. It is based on the GPT-3.5 architecture, which stands for "Generative Pre-trained Transformer 3.5". With the OpenAI API~\cite{openaiapi} service, we have access to various models including \texttt{gpt-3.5-turbo} which is the most capable GPT-3.5 model. 

Let us now consider an XML parser, libxml2, which is a software library for parsing XML documents that have a hierarchical structure and can conceptually be interpreted as an XML tree. To explore program states deep in the execution path, test inputs must be in a valid format because a broken-format input would be rejected early at the program's entry point. In other words, the program execution process can be divided into two stages: 1) the format parsing stage where all inputs can reach; and 2) the post-parsing stage which is deeper in the program path and only inputs conform to the syntax rules can reach. To fuzz such applications, the testcase should comply with the input format (\ie realistic input) expected by the program under testing (PUT) such that the execution passes format check and exercises core logic.

We now test the GPT model's capability to generate structured data with XML object as an example. A valid XML object should conform to several syntax rules~\cite{xmlrules}: 1) All XML elements must have a closing tag; 2) XML tags are case sensitive; 3) All XML elements must be properly nested; 4) All XML documents must have a root element; 5) Attribute values must always be quoted. 

We prompt \texttt{gpt-3.5-turbo} model to generate variations that are similar to but differ from a given sample input. According to the result in ~\autoref{tab:gptsay}, \texttt{gpt-3.5-turbo} is able to generate an XML document that differs from the sample input but is still of valid syntax. We include two response examples from ChatGPT. Example 1 retains the structure of the sample input as well as the set of tags, the only changes are with the text within pairs of tags. In example 2, the tags differ from the sample input but they still conform to the XML syntax rules.

Such mutation is hard to achieve for a mutation-based approach like AFL++. This is because AFL++ performs byte-level mutation which can easily break the aforementioned syntax rules and thus result in an ill-formatted testcase. Intuitively, it takes several rounds of byte-level mutations to transform the sample input into a seed like the response from ChatGPT. However, it is virtually impossible as each round of mutation is random. Even with dictionary mutation, fuzzer cannot align a valid tag with dictionary entries to mutate the tag as a whole. It will still break the syntax by dropping a dictionary entry in the middle of a tag. However, in order to explore all code regions of the target program like libxml2, both broken-format and valid XML documents are needed in order to expose bugs during and after the parsing stage. Therefore we propose to incorporate ChatGPT as an additional mutator for greybox fuzzers like AFL++. 

%% file: 3_method.tex
\section{Design and Implementation}

\begin{figure*}[t]
    \centering
    \includegraphics[width=0.80\linewidth]{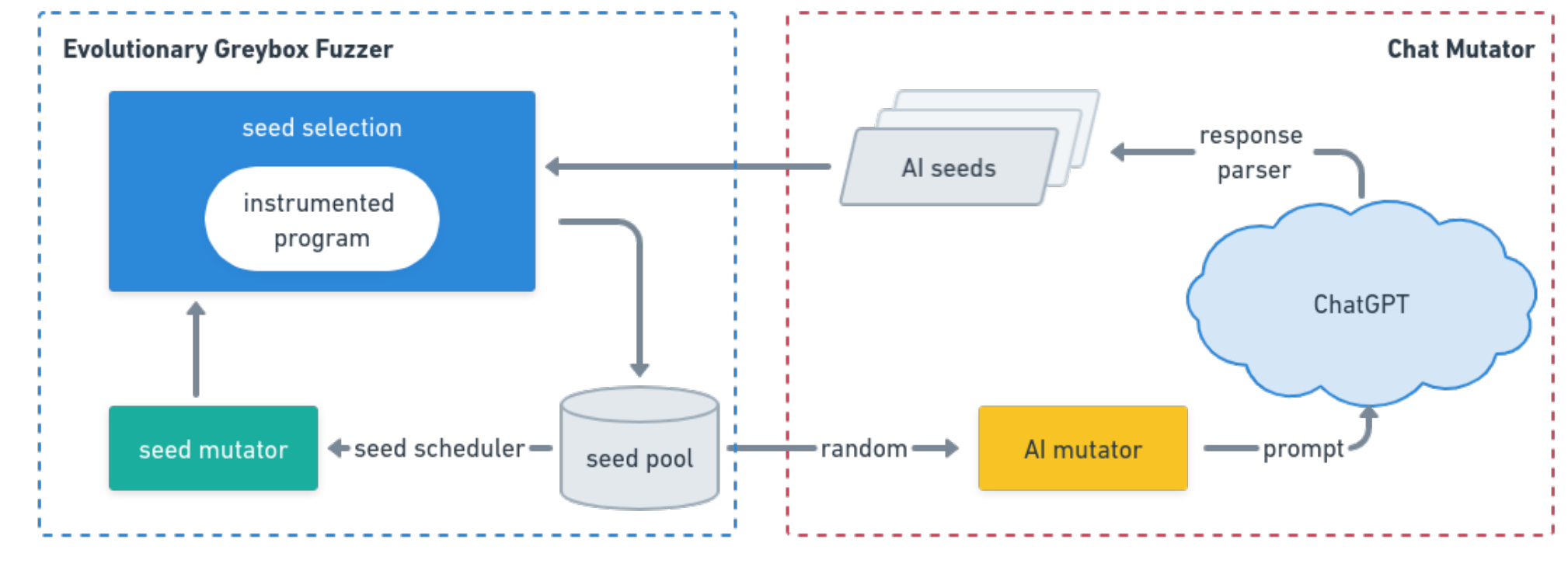}
    \vspace{-0.15in}
    \caption{\sys Overview}
    \label{fig:overview}
    \vspace{-0.2in}
\end{figure*}

In this section, we introduce \sys, a generative AI augmented greybox fuzzer implemented atop AFL++. 
As is shown in ~\autoref{fig:overview}, \sys comprises two major components: 1) an evolutionary greybox fuzzer which is the AFL++, and 2) a Chat Mutator, which is an AI-empowered seed mutator. 

At the beginning of the testing process, the Chat Mutator randomly picks one seed from the fuzzer's seed pool as sample input to prompt the ChatGPT model endpoint for variations. The response is then parsed to dump each variation as a seed stored in the AI seed queue. The greybox fuzzer counterpart of \sys can be any greybox fuzzer which will pick one seed from the seed pool for mutation and evaluate the mutated seeds to retain those that trigger new code region of the PUT. The seeds generated from the Chat Mutator will be periodically scanned to import the new-coverage seeds into the fuzzer's seed pool.

\subsection{Hyper-Parameter Selection}
The performance of \sys is dependent on the quality of the seeds generated from both counterparts. In order to find the best configuration of Chat Mutator without the influence of the greybox counterpart, we replace AFL++ with \texttt{afl-cmin} which will filter the seeds generated from Chat Mutator to filter the seeds by their edge coverage. We discuss five primary hyper-parameters of ChatGPT in this section.

\medskip\noindent\textit{\textbf{Hyper-parameters.}} Firstly, given the various models with different capabilities, we discuss which model is more suitable for our goal (section \ref{sec:models}) as well as provide the prompt templates for the models (section \ref{sec:prompt}). Secondly, as efficiency is a crucial factor for building a greybox fuzzer, we evaluate parameter choice of \texttt{max\_tokens} (section \ref{sec:mtoken}) and \texttt{n} (section \ref{sec:nchoi}) which are the two hyper-parameters highly associated with the model query latency. Thirdly, in order to produce high-quality seeds that improve code coverage, we find the optimal value choice for \texttt{temperature} (section \ref{sec:temper}). Last but not least, we find the necessary information to include in the prompt for generating better seeds (section \ref{sec:promfix}).

\medskip\noindent\textit{\textbf{Benchmarks.}} To avoid over-fitting, we conduct experiments on four programs: \texttt{jq}, \texttt{php}, \texttt{mujs} and \texttt{xml}. Note that \texttt{mujs} takes JavaScript source code as input while the rest three targets take various formatted data files (JSON, PHP, XML) as input. We evaluate the edge coverage growth (evaluated by afl-cmin tool shipped in AFL++ 4.03a), testcase unique ratio (validated by the md5sum checksum), and syntactically valid ratio.

\subsubsection{Model Endpoint Choice}\label{sec:models}

Various model endpoints are available in the OpenAI API depending on the type of input and the nature of the task to complete. In this paper, we focus on mutating sample input in the form of text-based data. Therefore, we have two model endpoints for evaluation: 1) the chat model and 2) the completion model. 

\medskip
\noindent\textit{\textbf{Chat Model.}} Given a list of messages describing a conversation, the chat model will return a response. Models under this category are \texttt{gpt-4, gpt-4-0314, gpt-4-32k, gpt-4-32k-0314, gpt-3.5-\ 
turbo} and \texttt{gpt-3.5-turbo-0301}. The most capable and stable model is \texttt{gpt-3.5-turbo}. Therefore we evaluate this particular model in the following discussion and use tag \textit{CT} to denote it. Note that at present, the \texttt{gpt-3.5-turbo} model charges \$0.0002 per 1000 tokens.

\medskip\noindent\textit{\textbf{Completion Model.}} With the completion model, we can provide a prompt for the model to return one or more predicated completions for it. Under this category we have  \texttt{text-davinci-003, text-davinci-\ 
002, text-curie-001, text-babbage-001} and \texttt{text-ada-001}. We conducted a small-scale test to check the latency of these models as well as the ratio of valid seeds generated. We choose \texttt{text-curie-\ 
001} as our completion model in the following of evaluation. This model is denoted as \textit{CP}. Compared to the \textit{CT} model, \textit{CP} is of higher efficiency and is more expensive. At present, the \texttt{text-curie-001} model charge \$0.0020 for every 1000 tokens. 

\begin{table*}[tbh]
\centering
\caption{Prompt Templates}
\label{tab:prompts}
\resizebox{\textwidth}{!}
{\small %
\begin{tabular}{ccccl}
\hline
\multirow{2}{*}{\textbf{Config}} &
  \multicolumn{2}{c}{\textbf{Prompt Info.}} &
  \multirow{2}{*}{\textbf{Model}} &
  \multicolumn{1}{c}{\multirow{2}{*}{\textbf{Prompt Template}}} \\ \cline{2-3}
 &
  \textbf{Sample Input} &
  \textbf{Format} &
   &
  \multicolumn{1}{c}{} \\ \hline
\multirow{3}{*}{\textbf{AI}} &
  \multirow{3}{*}{\cmark} &
  \multirow{3}{*}{\cmark} &
  CT &
  \begin{tabular}[c]{@{}l@{}}System: "You are a \textless{}format\textgreater file generator"\\ User: "Here is an example \textless{}format\textgreater file, generate another one." + \textless{}sample input\textgreater{}\end{tabular} \\ \cline{4-5} 
 &
   &
   &
  \multirow{2}{*}{CP} &
  \multirow{2}{*}{\textless{}sample input\textgreater + "And here is another \textless{}format\textgreater file like above: "} \\
 &
   &
   &
   &
   \\ \hline
\multirow{3}{*}{\textbf{AI\_noINPUT}} &
  \multirow{3}{*}{\xmark} &
  \multirow{3}{*}{\cmark} &
  CT &
  \begin{tabular}[c]{@{}l@{}}System: "You are a \textless{}format\textgreater file generator"\\ User: "Generate a \textless{}format\textgreater file."\end{tabular} \\ \cline{4-5} 
 &
   &
   &
  \multirow{2}{*}{CP} &
  \multirow{2}{*}{"Here is a \textless{}format\textgreater file: "} \\
 &
   &
   &
   &
   \\ \hline
\multirow{3}{*}{\textbf{AI\_noFORM}} &
  \multirow{3}{*}{\cmark} &
  \multirow{3}{*}{\xmark} &
  CT &
  \begin{tabular}[c]{@{}l@{}}System: "You are a file generator"\\ User: "Here is an example file, generate another one." + \textless{}sample input\textgreater{}\end{tabular} \\ \cline{4-5} 
 &
   &
   &
  \multirow{2}{*}{CP} &
  \multirow{2}{*}{\textless{}sample input\textgreater + "And here is another one like above: "} \\
 &
   &
   &
   &
   \\ \hline
\end{tabular}%
}
\vspace{-0.2in}
\end{table*}

\subsubsection{Prompt Design}\label{sec:prompt}

The design of the prompt words for querying the ChatGPT model is crucial to the response quality. Our goal is to improve greybox fuzzing by incorporating a generative LLM model as an additional mutator to generate more realistic inputs. Therefore, we instruct the model to generate variations of a sample input in a particular format. In order to evaluate the impact of different prompts over the generated testcases, we evaluate three different configurations of the prompt words for the two model endpoints, as is shown in ~\autoref{tab:prompts}. Basically, we need to include either a sample input or the expected format of the new testcases. Intuitively, the \textbf{AI} model with both sample input and file format known is the most informed configuration. We added the configuration \textbf{AI\_noINPUT} in order to understand if the sample input is important to generate better seeds for greybox fuzzer; as well as the configuration ~\textbf{AI\_noFORM} in order to see if providing the expected format of the input is necessary to generate format-conforming testcases.

\subsubsection{\texttt{max\_tokens} - Maximum number of tokens}\label{sec:mtoken}

This hyper-parameter can be an arbitrary positive integer value and it denotes the maximum length of the model response. With a smaller value, the generated testcase could be cut off short and result in a broken format. With a larger \texttt{max\_tokens} value, the model can take more time to finish the request and generate the response. However, well-formatted testcases do not necessarily have a fixed length. For data formats like JSON, PHP, JavaScript, XML, a valid seed can be of various lengths. A larger \texttt{max\_tokens} value is better for generating more valid seeds. However, with greybox fuzzing, it is important to generate as many mutations as possible within a short amount of time. Therefore, we want to strike a balance between efficiency and seed quality.

\begin{figure}[h]
    \centering
    \includegraphics[width=.7\linewidth]{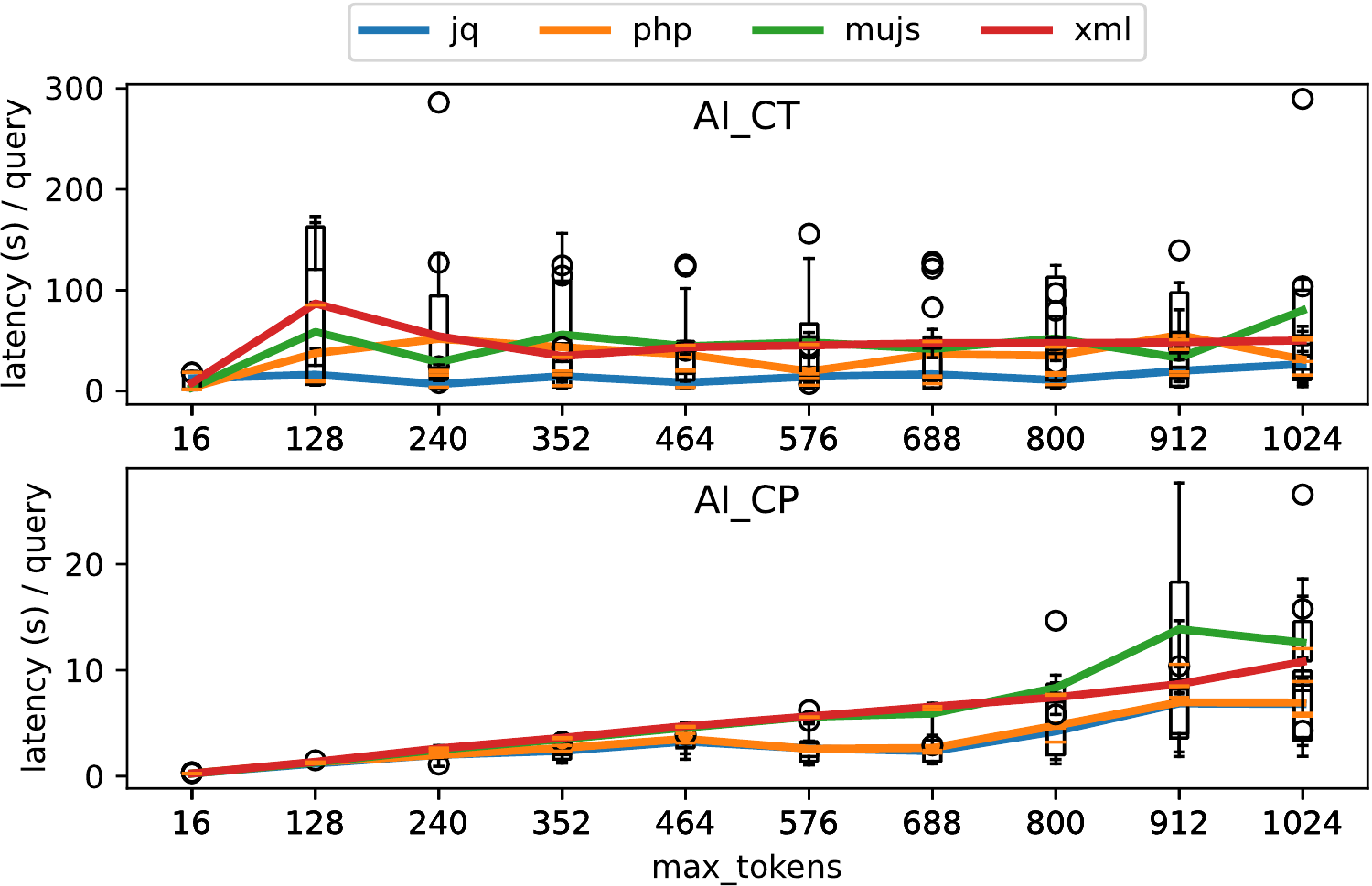}
    \caption{Model Latency and \texttt{max\_tokens}}
    \label{fig:mlate}
    \vspace{-0.1in}
\end{figure}

To find the best configuration of this hyper-parameter, we investigate the model latency for different \texttt{max\_tokens} values ranging from 16 (default value) to 1024 for \textbf{AI} configuration with both chat model (\ie \textit{AI\_CT}) and completion model (\ie \textit{AI\_CP}) respectively. For each program, we prompt the corresponding chat mutator with the same seed and record the latency. In order to reduce the randomness, we repeat each query for 10 times and draw the box plot as shown in ~\autoref{fig:mlate}. We also plot the average model latency for each \texttt{max\_tokens} value choice per program. 

\begin{table}[h]
\centering
\vspace{0.01in}
\caption{Sample Input Length by Token Count}
\label{tab:inlen}
{\small %
\begin{tabular}{ccccc}
\hline
\textbf{Program}       & \textbf{jq}              & \textbf{php}            & \textbf{mujs}                  & \textbf{xml}            \\ \hline
\textbf{Format}        & \multicolumn{1}{l}{JSON} & \multicolumn{1}{l}{PHP} & \multicolumn{1}{l}{JavaScript} & \multicolumn{1}{l}{XML} \\
\textbf{Sample Length} & 14                       & 100                     & 129                            & 450                     \\ \hline
\end{tabular}%
}
\vspace{-0.12in}
\end{table}

For each program, we use one syntactically valid seed for this experiment. We evaluate the length\footnote{https://platform.openai.com/tokenizer} of these sample inputs and report them along with the data format in ~\autoref{tab:inlen}. With model \textit{AI\_CP}, it is clear that as the value of \texttt{max\_tokens} increases, the model latency grows in general for the four tested programs. With  model \textit{AI\_CT}, the trend is not clear. We pick the largest power of 2 with a model latency lower than 5s per query. For the rest of the discussion, we let \texttt{max\_tokens} take the value 256.

\subsubsection{\texttt{n} - Completion choices number}\label{sec:nchoi}

This hyper-parameter determines how many completion choices to generate for each model query. The value of \texttt{n} is an integer between 1 and 128. Intuitively, the turnaround time for each query will be extended with a large \texttt{n} value. Furthermore, there is a context length limitation for each model query. The context includes both prompt/input and response/output. As more response choices are generated, if the context length reaches the limitation, no more response choices will be generated. To find the optimal value choice for \texttt{n}, we conduct the following experiment:

\begin{figure}[!htb]
   \begin{minipage}{0.49\textwidth}
     \centering
     \includegraphics[width=\linewidth]{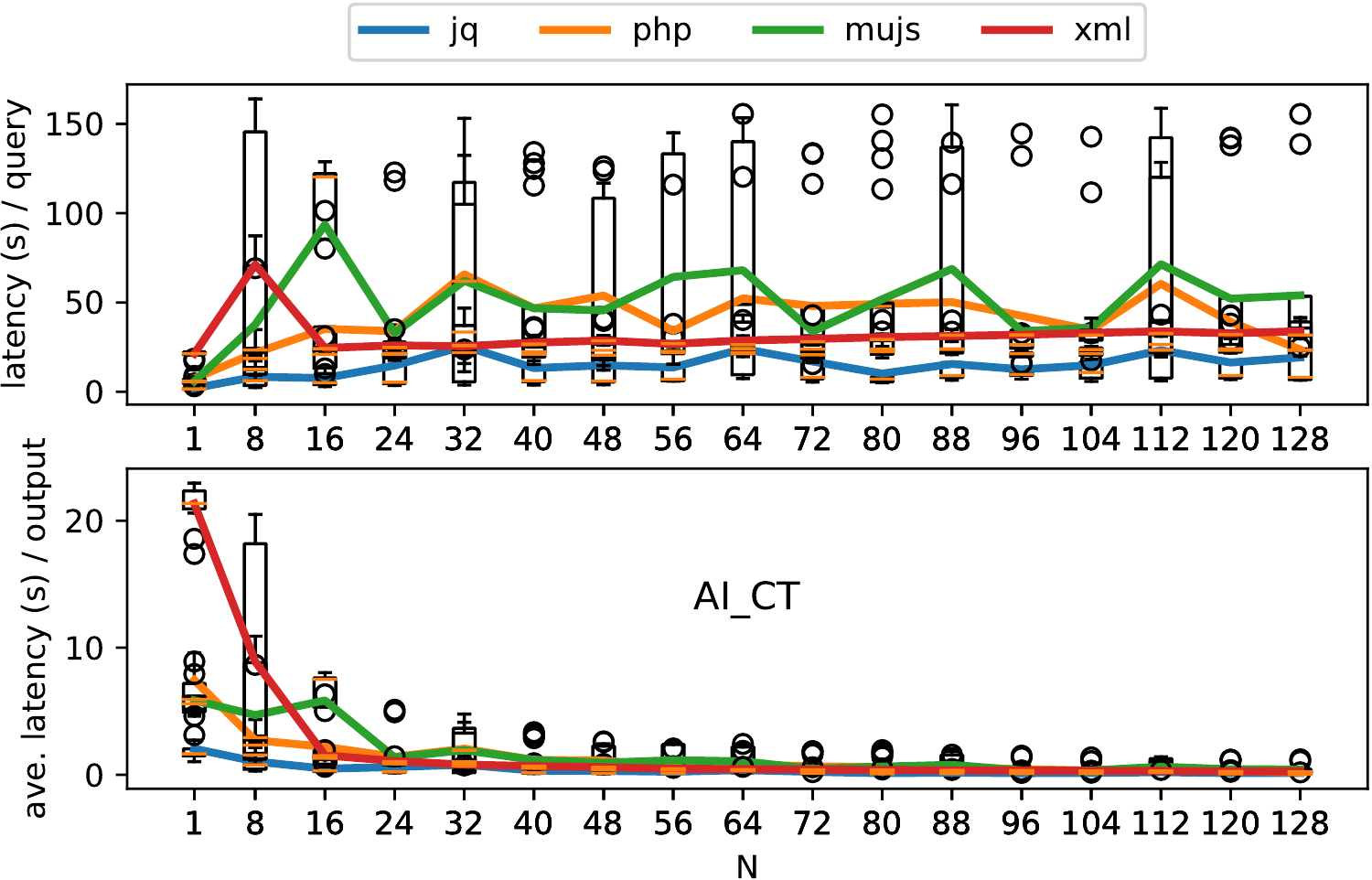}
     \caption{Model Latency and \texttt{n} for \textit{CT} Endpoint}
    \label{fig:CTN}
   \end{minipage}\hfill
   \begin{minipage}{0.49\textwidth}
     \centering
     \includegraphics[width=\linewidth]{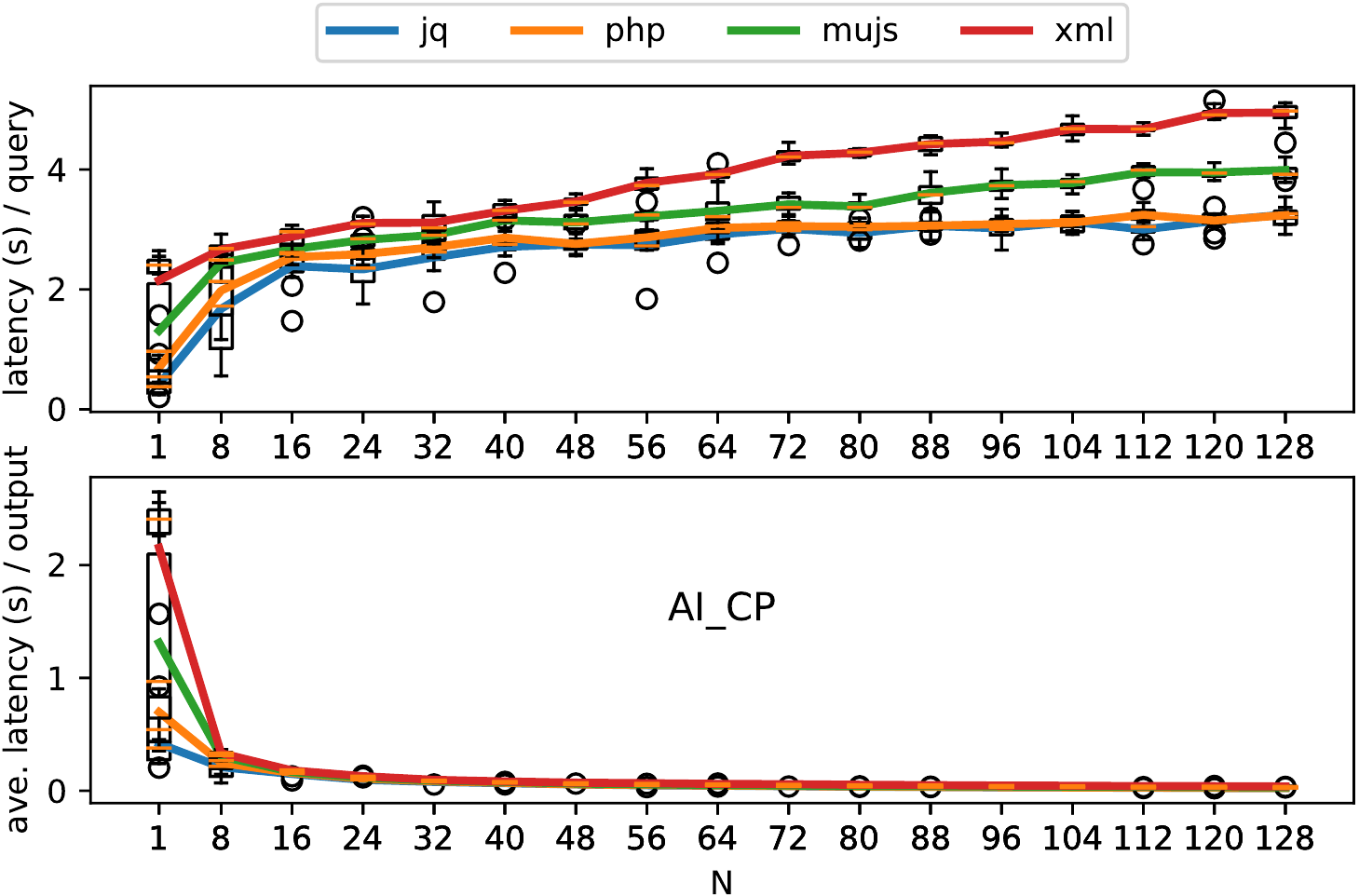}
     \caption{Model Latency and \texttt{n} for \textit{CP} Endpoint}
    \label{fig:CPN}
   \end{minipage}
   \vspace{-0.1in}
\end{figure}



For each program, we evaluate the latency for each model query as well as for each generated testcase on average with different \textbf{AI} model endpoints (\ie \textit{CT} and \textit{CP}) with different values of \texttt{n} from 1 to 128. Each configuration is repeated 10 times to reduce randomness. The same set of sample input from ~\autoref{tab:inlen} is used for this experiment. The result is shown in ~\autoref{fig:CTN} and ~\autoref{fig:CPN} for model \textit{AI\_CT} and \textit{AI\_CP} respectively. 

According to the result, as the value of \texttt{n} grows, the cost per query is increasing which means the time spent per input seed becomes higher, while the average cost for generating one testcase is decreasing, \ie higher efficiency. To advance the fuzzing progress and find new coverage efficiently, the chat mutator should have a short turnaround time to generate variations for more seeds in the fuzzer queue shortly. For the following discussions, we let \texttt{n} take the value 20.

\subsubsection{\texttt{temperature} - Sampling Temperature}\label{sec:temper}

This hyper-parameter denotes the value of the sampling temperature for querying the model. It takes a floating point value between 0 and 2 with a default value of 1. With a lower \texttt{temperature} value, the model is considered more focused and deterministic while a higher value will make the output more random. With a low \texttt{temperature} value, the model will give a response that is similar, sometimes even identical to the sample input. When the \texttt{temperature} value is high, the output becomes too random that the format of the output does not conform to the syntax rules anymore. In order to find the optimal \texttt{temperature} value choice, we perform the following experiments:

For each target program, we allow both models, \textit{AI\_CT} and \textit{AI\_CP}, to run continuously for two hours. For each model, we configure the sampling temperature to take values between 0 and 2 with a stepsize of 0.25. At the end of each trial, we evaluate the quality of the generated seeds with three metrics: 1) seed unique ratio, 2) seed valid ratio, and 3) code coverage. With the result, we discuss how to decide the best sampling temperature. 

\begin{figure}[!htb]
   \begin{minipage}{0.48\textwidth}
     \centering
     \includegraphics[width=\linewidth]{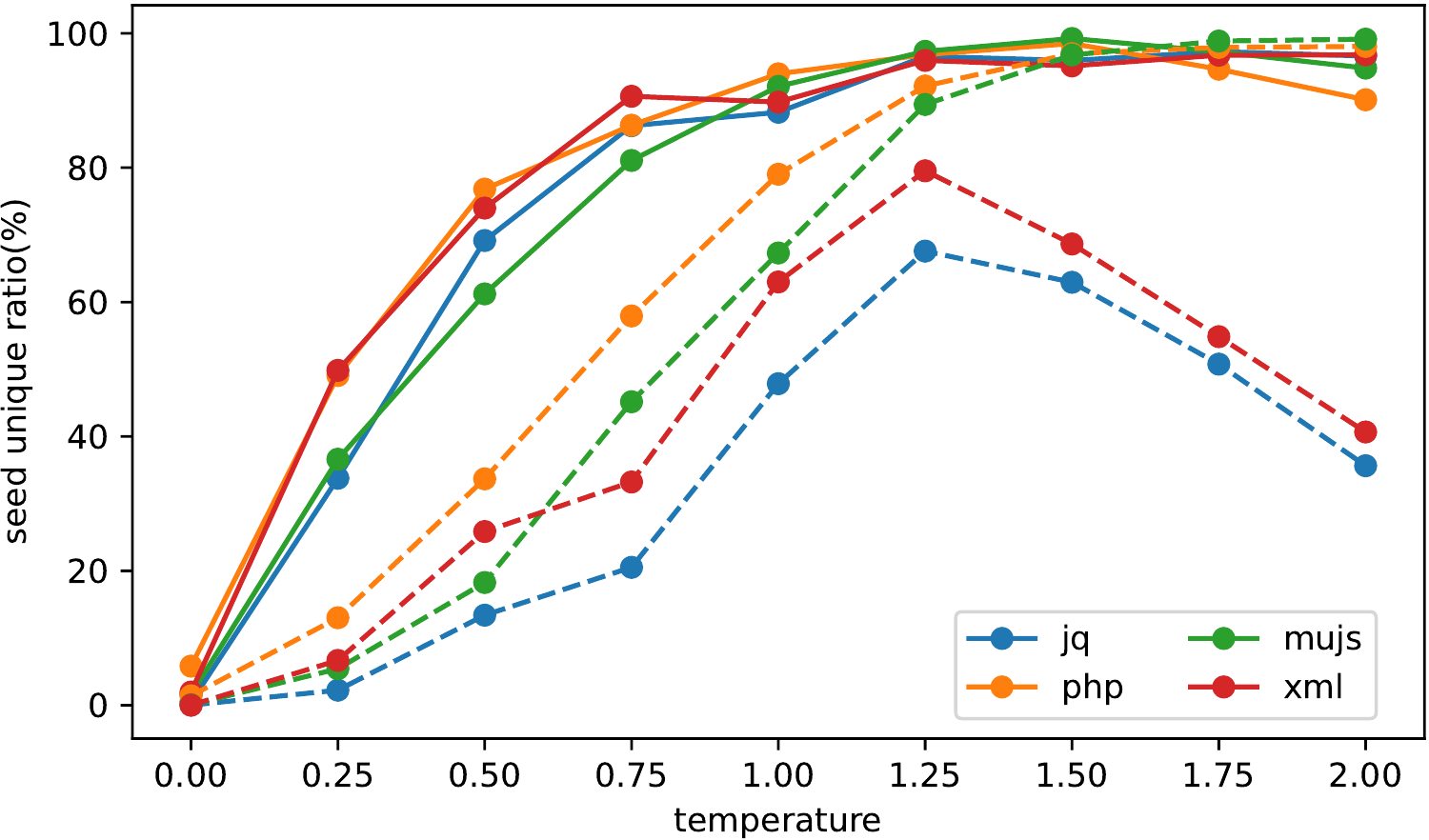}
     \caption{Seed unique ratio of all generated seeds. Note that the result of \textit{AI\_CT} is in a solid line while that of \textit{AI\_CP} is in a dashed line.}
    \label{fig:temp-uniq}
   \end{minipage}\hfill
   \begin{minipage}{0.48\textwidth}
     \centering
     \includegraphics[width=\linewidth]{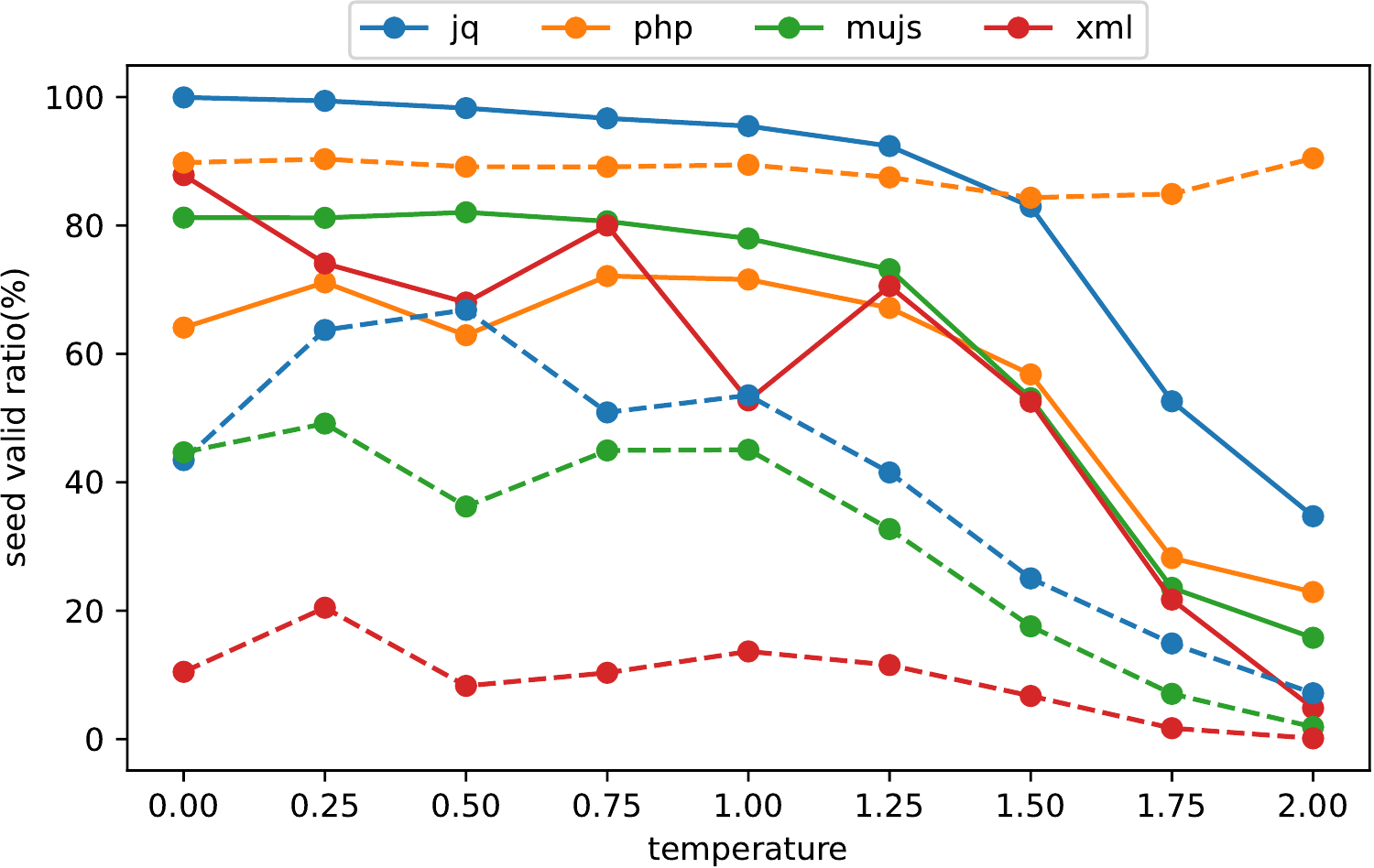}
     \caption{Seed valid ratio of all generated seeds. Note that result of \textit{AI\_CT} is in solid line while that of \textit{AI\_CP} is in dash line.}
    \label{fig:temp-valid}
   \end{minipage}
   \vspace{-0.2in}
\end{figure}

\medskip\noindent\textit{\textbf{Seed unique ratio. }} As is shown in ~\autoref{fig:temp-uniq}, for \textit{AI\_CT} model the ratio of unique seeds increases as temperature grows. When the sampling temperature is lower than 0.25, more than half of the generated seeds are duplicates. In general, with a higher sampling temperature, we can generate more unique seeds with \textit{AI\_CT} model. As for \textit{AI\_CP}, the seed unique ratio increases as sampling temperature grows from 0 to 1.25 and drops when the temperature is higher than 1.25 for program \texttt{jq} and \texttt{xml}. We find that more ill-formatted seeds that are only a couple of bytes long are generated within this temperature range. It is worth mentioning that seeds generated from \textit{AI\_CT} model are of higher unique ratio than that of \textit{AI\_CP} model. 

\medskip\noindent\textit{\textbf{Seed valid ratio. }} We also evaluate the ratio of syntactically valid seeds out of all the generated seeds for each configuration under different sampling temperature settings. According to the result in ~\autoref{fig:temp-valid}, the ratio of valid seeds decreases as the sampling temperature becomes higher. The reason is that, with a higher sampling temperature, the output of the model is more random and therefore more likely to break the syntax rules. It is worth mentioning that, \textit{AI\_CT} model has a higher seed valid ratio than \textit{AI\_CP} model for \texttt{jq, mujs, xml} while lower for \texttt{php}.

\begin{figure}[tbh]
    \centering
    \includegraphics[width=.7\linewidth]{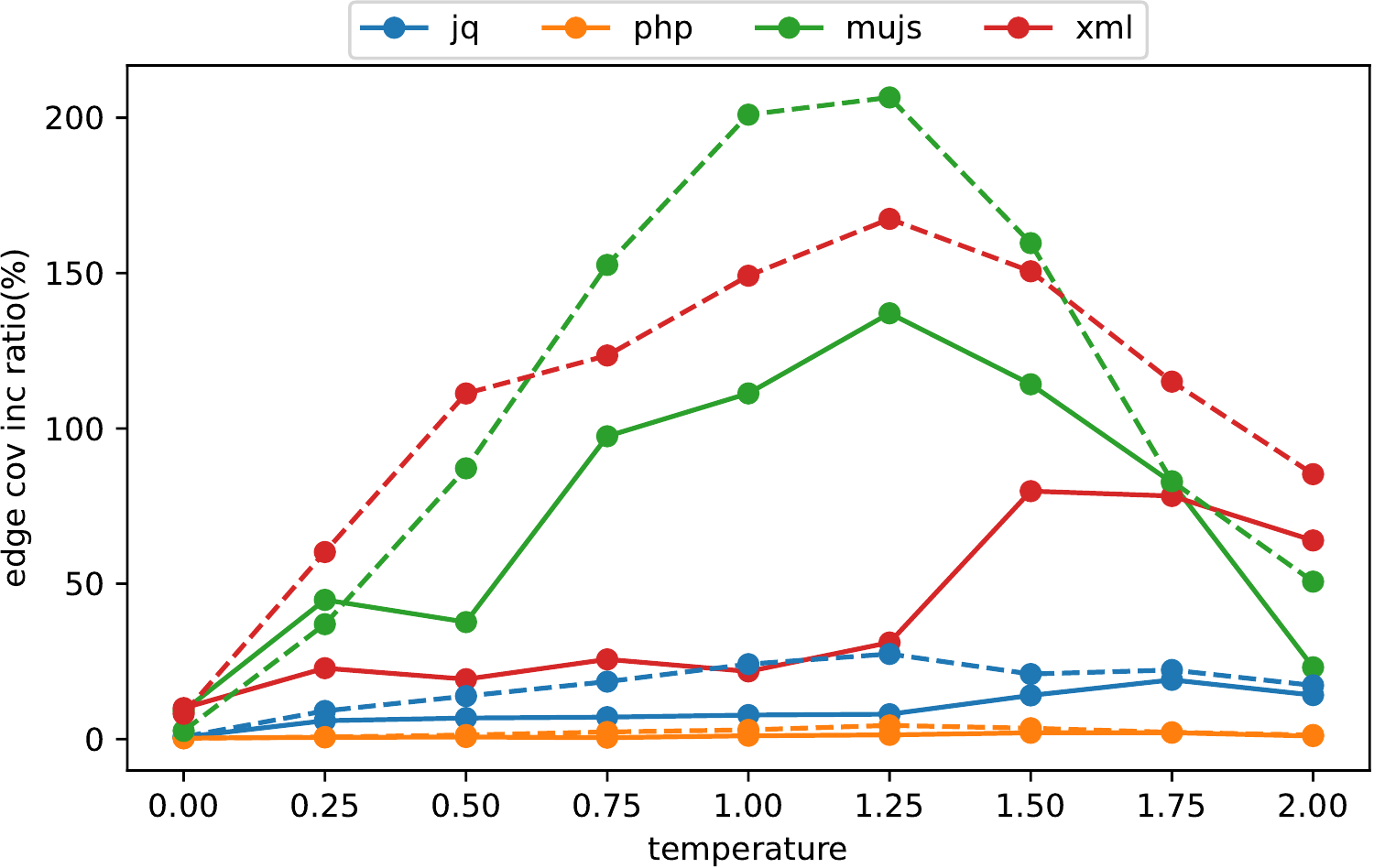}
    \caption{Code Coverage Improvement Over Initial Corpus}
    \label{fig:covinc}
\end{figure}

\medskip\noindent\textit{\textbf{Coverage improvement. }} For each model, we evaluate different sampling temperatures from 0 to 2 with a stepsize of 0.25. For each configuration, nine different \texttt{temperature} values are assessed. At the end of the 2h fuzzing trial, we collect the edge coverage of all generated seeds and compute the improvement ratio over the initial corpus and present the result in ~\autoref{fig:covinc}. We further investigate the statistical ranking of the code coverage improvement ratio on the four tested programs for different sampling temperatures and present the result in ~\autoref{tab:starank}. 

According to the result, \textit{AI\_CT} has the highest code coverage improvement with \texttt{temperature} 1.50 while \textit{AI\_CP} performs the best with \texttt{temperature} evaluating to 1.25. With a lower sampling temperature, the model generates a large amount of duplicates therefore the code coverage improvement is low. With a high sampling temperature, the model generates more syntactically invalid seeds, and that defeats the purpose of leveraging LLM for generating syntactically valid seeds. 

\begin{table}[h]
\centering
\caption{Statistical Ranking by Coverage Improvement Ratio}
\label{tab:starank}
{%
\begin{tabular}{cccccccccc}
\hline
\multirow{2}{*}{\textbf{Model}} & \multicolumn{9}{c}{\textbf{\texttt{temperature}}}                                     \\ \cline{2-10} 
                                & 0.00 & 0.25 & 0.50 & 0.75 & 1.00 & 1.25         & 1.50         & 1.75 & 2.00 \\ \hline
AI\_CT                          & 9.0  & 7.0  & 7.0  & 5.8  & 4.5  & 3.0          & \textbf{2.0} & 2.3  & 4.5  \\
AI\_CP                          & 9.0  & 8.0  & 6.3  & 4.3  & 2.5  & \textbf{1.0} & 2.8          & 4.8  & 6.5  \\ \hline
\end{tabular}%
}
\vspace{-0.15in}
\end{table}

When comparing the results between \textit{AI\_CT} and \textit{AI\_CP}, the seeds generated from \textit{AI\_CT} are of better quality as the seed valid ratio and unique ratio are both higher than that of \textit{AI\_CP}. However, \textit{AI\_CT} outperforms \textit{AI\_CP} in terms of the coverage improvement ratio. This is because \textit{AI\_CT} model is of much higher efficiency than \textit{AI\_CP}. Within the 2h fuzzing trial time, \textit{AI\_CP} model generates 15.91 $\sim$ 99.31$\times$ more seeds than \textit{AI\_CT}.

\subsubsection{Prompt Ablation Study}\label{sec:promfix}

\medskip
\noindent\textit{\textbf{Sample Input. }} Here we discuss the necessity for including a sample input in the model prompt. For each program, we evaluate the configuration \textbf{AI} and \textbf{AI\_noINPUT} for both the chat model (\ie CT) and completion model (\ie CP) and evaluate the edge coverage growth in eight hours.

\begin{figure}[!htb]
   \begin{minipage}{0.48\textwidth}
     \centering
     \includegraphics[width=\linewidth]{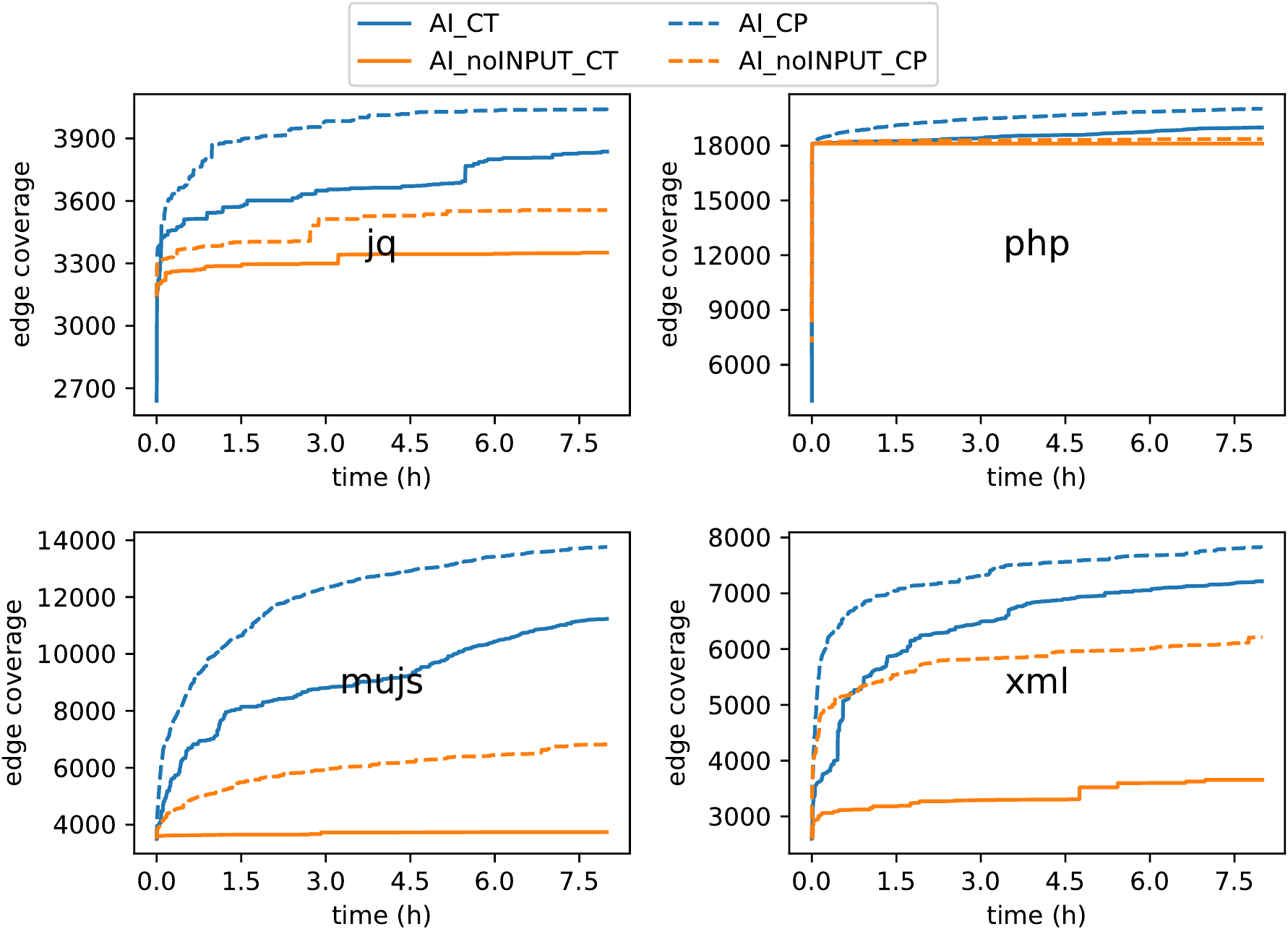}
    \caption{\textit{AI} vs. \textit{AI\_noINPUT}}\label{fig:noinput}
   \end{minipage}\hfill
   \begin{minipage}{0.48\textwidth}
     \centering
     \includegraphics[width=\linewidth]{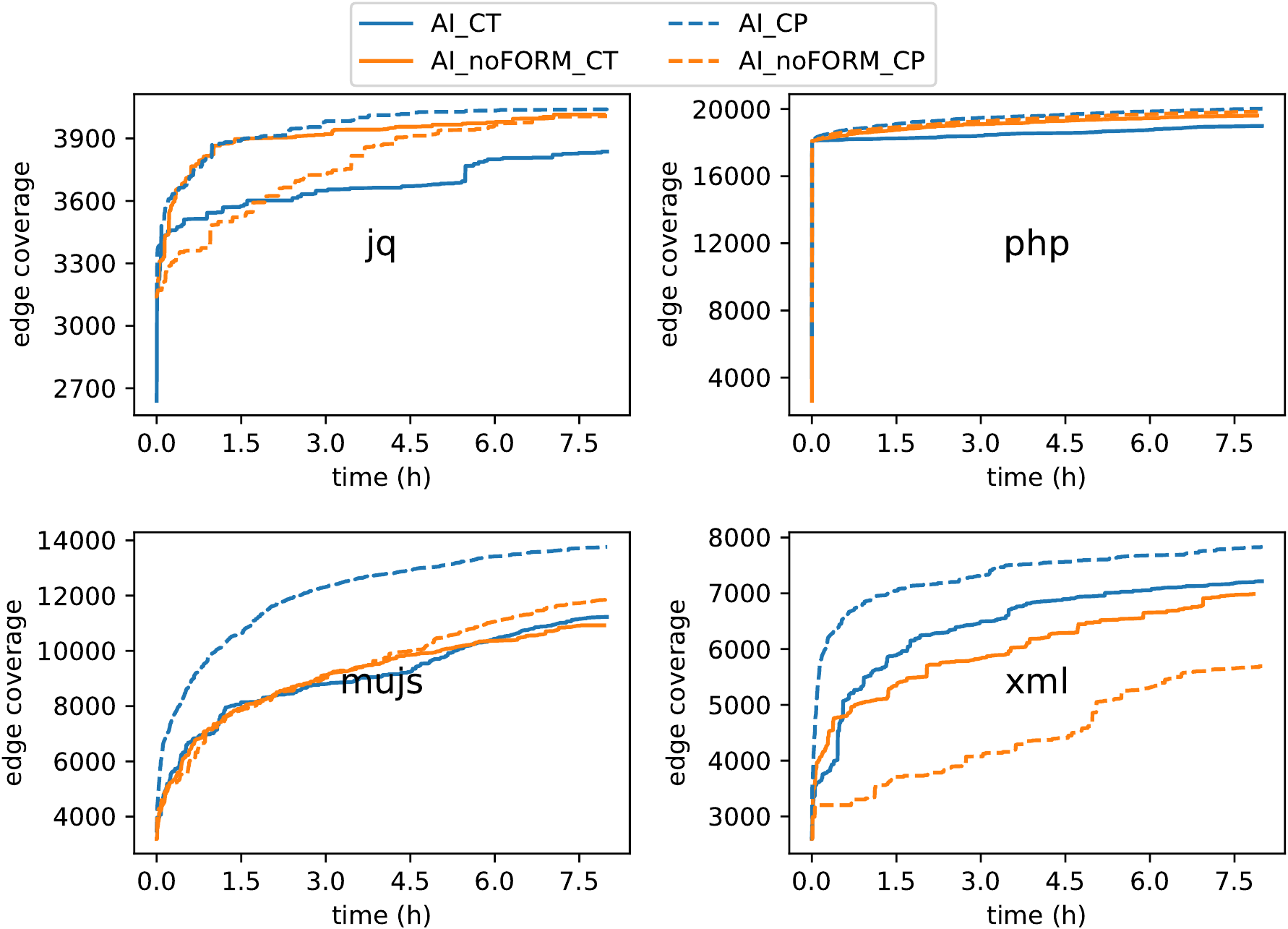}
    \caption{\textit{AI} vs. \textit{AI\_noFORM}}
    \label{fig:noformat}
   \end{minipage}
\end{figure}

According to the result shown in \autoref{fig:noinput}, across the four tested programs, \textbf{AI} consistently performs better than \textbf{AI\_noINPUT} where no sample input is included in the model prompt. The result indicates that sample input is crucial for the generative AI model to generate seeds that trigger new code coverage. In terms of model endpoints, the \textit{CP} model performs consistently better than the \textit{CT} model. We report the edge coverage at the end of the trial in \autoref{tab:promabla}. For the \textit{CT} model, the code coverage findings decrease on average by 15.98\% when the sample input is removed from the model prompt. For \textit{CP} model, removing sample input from the model prompt results in 22.88\% code coverage decreasing. 

\begin{table*}[th]
\centering
\caption{Prompt Ablation Study}
\label{tab:promabla}
{\small %
\begin{tabular}{c|rrrrr|rrrrr}
\hline
\multirow{3}{*}{\textbf{Program}} &
  \multicolumn{5}{c|}{\textbf{CT endpoint}} &
  \multicolumn{5}{c}{\textbf{CP endpoint}} \\ \cline{2-11} 
 &
  \multicolumn{1}{c|}{\textbf{AI}} &
  \multicolumn{2}{c|}{\textbf{AI\_noFORM}} &
  \multicolumn{2}{c|}{\textbf{AI\_noINPUT}} &
  \multicolumn{1}{c|}{\textbf{AI}} &
  \multicolumn{2}{c|}{\textbf{AI\_noFORM}} &
  \multicolumn{2}{c}{\textbf{AI\_noINPUT}} \\ \cline{2-11} 
 &
  \multicolumn{1}{c|}{cov} &
  \multicolumn{1}{c}{cov} &
  \multicolumn{1}{c|}{vs. AI} &
  \multicolumn{1}{c}{cov} &
  \multicolumn{1}{c|}{vs. AI} &
  \multicolumn{1}{c|}{cov} &
  \multicolumn{1}{c}{cov} &
  \multicolumn{1}{c|}{vs. AI} &
  \multicolumn{1}{c}{cov} &
  \multicolumn{1}{c}{vs. AI} \\ \hline
jq &
  \multicolumn{1}{r|}{3837} &
  4015 &
  \multicolumn{1}{r|}{+4.64\%} &
  3555 &
  -7.35\% &
  \multicolumn{1}{r|}{4043} &
  4015 &
  \multicolumn{1}{r|}{-0.69\%} &
  3555 &
  -12.07\% \\
php &
  \multicolumn{1}{r|}{18995} &
  19609 &
  \multicolumn{1}{r|}{+3.23\%} &
  18364 &
  -3.32\% &
  \multicolumn{1}{r|}{20021} &
  19609 &
  \multicolumn{1}{r|}{-2.06\%} &
  18364 &
  -8.28\% \\
mujs &
  \multicolumn{1}{r|}{11233} &
  10924 &
  \multicolumn{1}{r|}{-2.75\%} &
  6819 &
  -39.29\% &
  \multicolumn{1}{r|}{13763} &
  10924 &
  \multicolumn{1}{r|}{-20.63\%} &
  6819 &
  -50.45\% \\
xml &
  \multicolumn{1}{r|}{7217} &
  6988 &
  \multicolumn{1}{r|}{-3.17\%} &
  6209 &
  -13.97\% &
  \multicolumn{1}{r|}{7832} &
  6988 &
  \multicolumn{1}{r|}{-10.78\%} &
  6209 &
  -20.72\% \\ \hline
\textbf{Average} &
  \multicolumn{1}{l|}{\textbf{}} &
  \multicolumn{1}{l}{\textbf{}} &
  \multicolumn{1}{r|}{\textbf{+0.49\%}} &
  \multicolumn{1}{l}{\textbf{}} &
  \textbf{-15.98\%} &
  \multicolumn{1}{l|}{\textbf{}} &
  \multicolumn{1}{l}{\textbf{}} &
  \multicolumn{1}{r|}{\textbf{-8.54\%}} &
  \multicolumn{1}{l}{\textbf{}} &
  \textbf{-22.88\%} \\ \hline
\end{tabular}%
}
\end{table*}

\medskip
\noindent\textit{\textbf{Input Format. }} Here we discuss the necessity for indicating the format of the testcase in the model prompt. For each program, we evaluate the configuration \textbf{AI} against \textbf{AI\_noFORM} for both the \textit{CT} chat model and \textit{CP} completion model. The result for 8h continuous fuzzing result is presented in \autoref{fig:noformat}. 

According to the result, removing data format information from the model prompt for \textit{CT} configuration does not substantially affect the coverage finding, as the coverage difference compared with \textbf{AI} model is less than 5\% for all four programs. As for \textit{CP} configuration, the code coverage dropped 8.54\% on average across the four tested programs.

%% file: 4_results.tex
\section{Evaluation}

In this section, we evaluate the efficacy of our proposed approach by answering the following research questions:


\begin{itemize}
    \item \textbf{RQ1: Coverage Efficiency.} How much coverage improvement can \sys make for end-to-end greybox fuzzing?  
    \item \textbf{RQ2: Mutation Effectiveness. } Can \sys generate high-quality seeds for all programs with different input formats?
    \item \textbf{RQ3: Security Application.}  Is \sys more effective for detecting vulnerabilities? 
\end{itemize}

\subsection{Evaluation Plan}

\smallskip
\noindent\textit{\textbf{Baselines.}} To better answer the aforementioned research questions, we use the following baseline configurations listed in ~\autoref{tab:blines} and explain each baseline below:

\begin{table}[ht]
\centering
\caption{Baselines}
\label{tab:blines}
{\small %
\begin{tabular}{lcc}
\hline
\textbf{Baseline} & \textbf{Model Endpoint} & \textbf{Format Agnostic?} \\ \hline
AFL++           & -              & -               \\
\sys            & CP             & \xmark          \\
\sys-F          & CP             & \cmark          \\
\sys-C          & CT             & \xmark          \\
\sys-CF         & CT             & \cmark          \\ \hline
\end{tabular}%
}
\vspace{-0.2in}
\end{table}

\begin{itemize}
    \item \textbf{AFL++}. The original AFL++~\cite{fioraldi2020afl++} fuzzer in non-deterministic mode. This configuration has one AFL++ 4.03a instance and no chat mutator. 
    \item \textbf{\sys}. The full-fledged \sys model integrating \textit{AI\_CP} with AFL++. In this configuration, the chat mutator randomly picks one seed from the fuzzer queue as sample input to prompt the \textit{CP} model for new variations with the optimal parameter configurations. All seeds generated from the chat mutator are periodically inspected by AFL++ to import any seeds that trigger new edge coverage into the fuzzer queue. 
    \item \textbf{\sys-F}. This model integrates AFL++ with \textit{AI\_noFORM\_CP}. Compared with the full-fledged \sys, this baseline removes data format information from the model prompt. In other words, this is the format-agnostic setting of \sys. 
    \item \textbf{\sys-C}. This model substitutes the \textit{CP} model with \textit{CT} model in \sys. The \textit{CT} model has significantly higher latency than the \textit{CP} model but is able to generate fewer duplicates and more valid seeds. 
    \item \textbf{\sys-CF}. With this model, we remove the input format information from the model prompt of the chat mutator for the \sys-C model. This is the format-agnostic setting of the \sys-C model. 
\end{itemize}

\medskip
\noindent\textit{\textbf{Dataset.}} Currently, \sys supports target programs that take in text-based formatted inputs. We inspect 45 programs from three benchmarks heavily evaluated in precedent works: Unibench, Fuzzbench and LAVA-M and list 12 programs that \sys can generate realistic inputs for in ~\autoref{tab:benchs}. Specifically, we classify the 12 targets into three categories based on the format of the input seeds: 1) formatted data file; 2) source code in different programming languages; and 3) text with no explicit syntax rules. We further show the reasons why the other 33 programs are not included in the evaluation in ~\autoref{tab:badform}.

\begin{table}[htb]
\caption{Benchmarks} 
\centering
\label{tab:benchs}
{%
\begin{tabular}{c|lll}
\hline
\textbf{Type}                                                                   & \textbf{Program} & \textbf{Version}        & \textbf{Input Format} \\ \hline
\multirow{4}{*}{\begin{tabular}[c]{@{}c@{}}data \end{tabular}}  & jq               & jq-1.5                  & json                          \\
                                                                                & php              & php-fuzz-parser\_0dbedb & PHP                                  \\
                                                                                & xml              & libxml2-v2.9.2          & XML                                 \\
                                                                                & jsoncpp\_fuzzer  & jsoncpp                 & json                                 \\ \hline
\multirow{3}{*}{\begin{tabular}[c]{@{}c@{}}code \end{tabular}} & mujs             & mujs-1.0.2              & js                                   \\
                                                                                & ossfuzz          & sqlite3\_c78cbf2        & SQL                                \\
                                                                                & cflow            & cflow-1.6               & C                                  \\ 
                                                                                & lua              & lua\_dbdc74d            & lua                                  \\ \hline
\multirow{4}{*}{\begin{tabular}[c]{@{}c@{}}text \end{tabular}} & curl\_fuzzer\_http & curl\_fuzzer\_9a48d43  & HTTP response         \\
                                                                                & openssl\_x509    & openssl-3.0.7           & DER certificate                      \\
                                                                                & base64           & LAVA-M                  & .b64 file                          \\
                                                                                & md5sum           & LAVA-M                  & md5 checksum                      \\ \hline
\end{tabular}%
}
\vspace{-0.05in}
\end{table}

\begin{table}[ht]
\centering
\caption{Unsupported Targets}
\label{tab:badform}
{%
\begin{tabular}{lll}
\hline
\textbf{Benchmark}          & \textbf{Targets}               & \textbf{Format} \\ \hline
\multirow{16}{*}{Unibench~\cite{li2021unifuzz}}  & exiv2                          & image                       \\
                            & gdk-pixbuf-pixdata             & image                       \\
                            & imginfo                        & image                       \\
                            & jhead                          & image                       \\
                            & tiffsplit                      & image                       \\
                            & lame                           & audio                       \\
                            & mp3gain                        & audio                       \\
                            & wav2swf                        & audio                       \\
                            & ffmpeg                         & video                       \\
                            & flvmeta                        & video                       \\
                            & mp42aac                        & video                       \\
                            & nm                             & binary                      \\
                            & objdump                        & binary                      \\
                            & tcpdump                        & network                     \\
                            & infotocap                      & terminfo file               \\
                            & pdftotext                      & pdf                         \\ \hline
\multirow{15}{*}{Fuzzbench~\cite{fuzzbench:misc}} & libjpeg-turbo-07-2017          & image                       \\
                            & libpng-1.2.56                  & image                       \\
                            & bloaty\_fuzz\_target           & ELF, Mach-O, etc.    \\
                            & freetype2-2017                 & TTF, OTF, WOFF              \\
                            & harfbuzz-1.3.2                 & TTF, OTF, TTC               \\
                            & lcms-2017-03-21                & ICC profile                 \\
                            & libpcap\_fuzz\_both            & PCAP                        \\
                            & mbedtls\_fuzz\_dtlsclient      & other                       \\
                            & openthread-2019-12-23          & other                       \\
                            & proj4-2017-08-14               & other                       \\
                            & re2-2014-12-09                 & other                       \\
                            & systemd\_fuzz-link-parser      & other                       \\
                            & vorbis-2017-12-11              & OGG                         \\
                            & woff2-2016-05-06               & WOFF                        \\
                            & zlib\_zlib\_uncompress\_fuzzer & Zlib compressed             \\ \hline
\multirow{2}{*}{LAVA-M~\cite{dolan2016lava}}     & who                            & utmp file                   \\
                            & uniq                           & other                       \\ \hline
\end{tabular}%
}
\vspace{-0.2in}
\end{table}

\begin{table*}[h]
\centering
\caption{Coverage Analysis}
\label{tab:covres}
\resizebox{\textwidth}{!}{%
\begin{tabular}{rrrrrrrrrr}
\hline
\multirow{2}{*}{\textbf{Program}} &
  \multicolumn{5}{c}{\textbf{8h Code Coverage}} &
  \multicolumn{4}{c}{\textbf{Improvement over AFL++}} \\ \cmidrule(lr){2-6} \cmidrule(lr){7-10} 
 &
  AFL++ &
  \sys &
  \sys-F &
  \sys-C &
  \sys-CF &
  \sys &
  \sys-F &
  \sys-C &
  \sys-CF \\ \hline
\multicolumn{1}{r|}{jq} &
  4045 &
  4109 &
  4122 &
  4101 &
  \multicolumn{1}{r|}{\textbf{4124}} &
  1.58\% &
  1.90\% &
  1.38\% &
  \textbf{1.95\%} \\
\multicolumn{1}{r|}{php} &
  22083 &
  \textbf{22507} &
  22308 &
  22388 &
  \multicolumn{1}{r|}{22389} &
  \textbf{1.92\%} &
  1.02\% &
  1.38\% &
  1.39\% \\
\multicolumn{1}{r|}{xml} &
  7687 &
  \textbf{9555} &
  9067 &
  9027 &
  \multicolumn{1}{r|}{9189} &
  \textbf{24.30\%} &
  17.95\% &
  17.43\% &
  19.54\% \\
\multicolumn{1}{r|}{jsoncpp\_fuzzer} &
  1236 &
  \textbf{1248} &
  1247 &
  \textbf{1248} &
  \multicolumn{1}{r|}{1250} &
  \textbf{0.97\%} &
  0.89\% &
  \textbf{0.97\%} &
  1.13\% \\ \hline
\multicolumn{1}{r|}{mujs} &
  9149 &
  \textbf{16171} &
  11982 &
  13194 &
  \multicolumn{1}{r|}{13023} &
  \textbf{76.75\%} &
  30.97\% &
  44.21\% &
  42.34\% \\
\multicolumn{1}{r|}{ossfuzz} &
  28880 &
  32751 &
  \textbf{33105} &
  32582 &
  \multicolumn{1}{r|}{32120} &
  13.40\% &
  \textbf{14.63\%} &
  12.82\% &
  11.22\% \\
\multicolumn{1}{r|}{cflow} &
  2499 &
  \textbf{2513} &
  2511 &
  2495 &
  \multicolumn{1}{r|}{2501} &
  \textbf{0.56\%} &
  0.48\% &
  -0.16\% &
  0.08\% \\
\multicolumn{1}{r|}{lua} &
  15213 &
  15530 &
  \textbf{15707} &
  15633 &
  \multicolumn{1}{r|}{15585} &
  \textbf{2.08\%} &
  3.25\% &
  2.76\% &
  2.45\% \\ \hline
\multicolumn{1}{r|}{curl} &
  13476 &
  13761 &
  \textbf{13762} &
  13687 &
  \multicolumn{1}{r|}{13699} &
  2.11\% &
  \textbf{2.12\%} &
  1.57\% &
  1.65\% \\
\multicolumn{1}{r|}{openssl\_x509} &
  18000 &
  17976 &
  18015 &
  \textbf{18016} &
  \multicolumn{1}{r|}{17970} &
  -0.13\% &
  0.08\% &
  \textbf{0.09\%} &
  -0.17\% \\
\multicolumn{1}{r|}{base64} &
  368 &
  368 &
  368 &
  368 &
  \multicolumn{1}{r|}{\textbf{372}} &
  0.00\% &
  0.00\% &
  0.00\% &
  \textbf{1.09\%} \\
\multicolumn{1}{r|}{md5sum} &
  482 &
  \textbf{525} &
  482 &
  \textbf{525} &
  \multicolumn{1}{r|}{482} &
  \textbf{8.92\%} &
  0.00\% &
  \textbf{8.92\%} &
  0.00\% \\ \hline
\multicolumn{1}{r|}{\textbf{Average}} &
  10259.8 &
  \textbf{11417.8} &
  11056.3 &
  11105.3 &
  \multicolumn{1}{r|}{11058.6} &
  \textbf{11.04\%} &
  6.11\% &
  7.61\% &
  6.89\% \\ \hline
\end{tabular}%
}
\end{table*}

\medskip
\noindent\textit{\textbf{Experiment Setup.}} All experiments were conducted on a workstation with two-socket, 48-core, 96-thread Intel Xeon Platinum 8168 processors. The workstation has 768G memory. The operating system is Ubuntu 18.04 with kernel 5.4.0. Note that each fuzzing trial is assigned one dedicated core to ensure fair comparison.

\subsection{RQ1: Coverage Efficiency}

\begin{figure*}[th]
    \centering
    \includegraphics[width=0.95\linewidth]{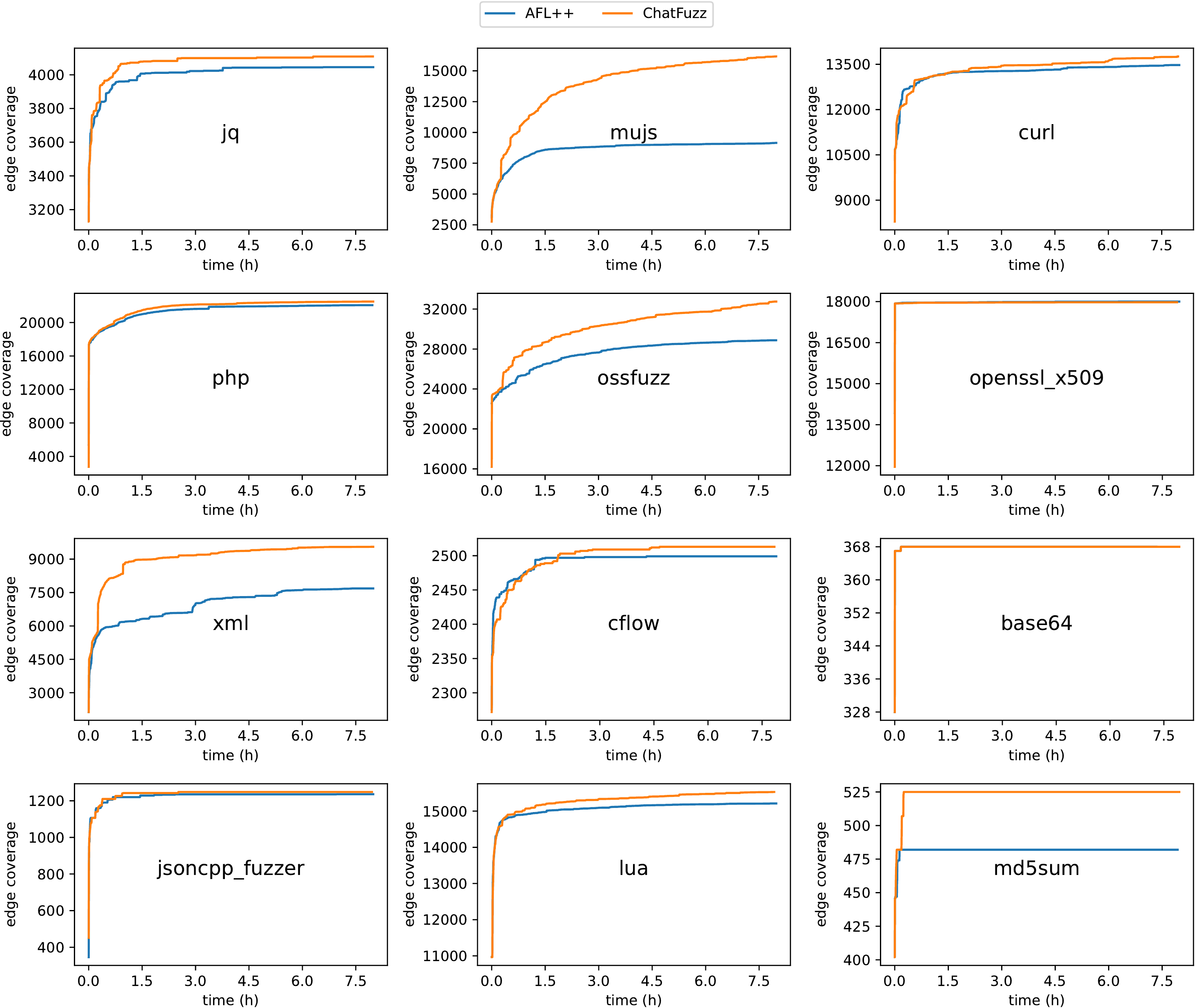}
    \caption{End-to-End Fuzzing Coverage Growth}
    \label{fig:end2end} 
    \vspace{-0.15in}
\end{figure*}

To demonstrate the coverage efficiency of \sys against AFL++ we measure the edge coverage for 8h continuous fuzzing campaigns for 12 programs. Result is shown in ~\autoref{fig:end2end}. The first column shows the result for four programs which take as input data files in the format of \texttt{JSON, PHP, XML} and \texttt{JSON}. The middle column shows the result for four programs which take program source code files (\texttt{JavaScript, SQL, C} and \texttt{lua}) as input. The last column shows the result for programs which take text-based files with no explicit syntax rules: HTTP response, DER certificate, .b64 file as well as md5 checksum.

Across the 12 target programs, \sys finds 11.04\% more edges than AFL++. Specifically, for program \texttt{xml, mujs, ossfuzz} and \texttt{md5sum}, \sys is able to find 24.30\%, 76.75\%, 13.40\% and 8.92\% more edge coverage than AFL++ correspondingly. The improvement over AFL++ is within 5\% on six programs: \texttt{jq, php, jsoncpp\_fuzzer, lua, curl} and \texttt{base64}. For program \texttt{cflow, openssl\_x509} and \texttt{base64} the difference is less than 1\%.

Moreover, we evaluate the coverage findings of \sys-F over the same set of targets to validate the potential of building a format-agnostic greybox fuzzer. For real-world program fuzzing, there are cases where the input format is not explicitly indicated. With powerful generative AI like ChatGPT, the model can potentially detect the format of the sample input and implicitly leverage this information for generating new mutations. We report the 8h code coverage of \sys-F in ~\autoref{tab:covres}. On average, it is able to find 6.11\% more coverage than AFL++. This indicates that, even without providing the expected format of the input, \sys is still able to improve greybox fuzzing. 

As precedent experiment in section \ref{sec:temper} suggests, the \textit{CT} endpoint generates less duplicates and is more likely to produce valid seeds than \textit{CP} endpoint. Therefore, we evaluate \sys-C and \sys-CF which are basically the \sys and \sys-F model with the \textit{CT} endpoint instead of \textit{CP} endpoint. We report the 8h fuzzing result in ~\autoref{tab:covres}. According to the result, \sys-C and \sys-CF find 7.61\% and 6.89\% more edges than AFL++. The result is close to the format-agnostic \sys fuzzer, \ie \sys-F.

The results demonstrate that \sys is of higher coverage efficiency than AFL++. 

\subsection{RQ2: Mutation Effectiveness}

In this subsection, we investigate the contribution of the generative AI-augmented seed mutator towards the code coverage. In ~\autoref{tab:seedimport}, we report the length of the fuzzer queue retained by AFL++ counterpart of the \sys system, along with the number of seeds imported from the chat mutator as well as the ratio of imported seeds over the fuzzer queue.

\begin{table*}[th]
\centering
\caption{Seed Analysis}
\label{tab:seedimport}
\resizebox{\textwidth}{!}{%
\begin{tabular}{rrrrrrrrrrrrr}
\hline
\multirow{3}{*}{\textbf{Program}} &
  \multicolumn{3}{c}{\textbf{\sys}} &
  \multicolumn{3}{c}{\textbf{\sys-F}} &
  \multicolumn{3}{c}{\textbf{\sys-C}} &
  \multicolumn{3}{c}{\textbf{\sys-CF}} \\ \cmidrule(lr){2-4} \cmidrule(lr){5-7} \cmidrule(lr){8-10} \cmidrule(lr){11-13}
 &
  \multicolumn{1}{c}{\multirow{2}{*}{FQ}} &
  \multicolumn{2}{c}{imported from AI} &
  \multicolumn{1}{c}{\multirow{2}{*}{FQ}} &
  \multicolumn{2}{c}{imported from AI} &
  \multicolumn{1}{c}{\multirow{2}{*}{FQ}} &
  \multicolumn{2}{c}{imported from AI} &
  \multicolumn{1}{c}{\multirow{2}{*}{FQ}} &
  \multicolumn{2}{c}{imported from AI} \\ \cline{3-4} \cline{6-7} \cline{9-10} \cline{12-13} 
 &
  \multicolumn{1}{c}{} &
  \multicolumn{1}{c}{seed count} &
  \multicolumn{1}{c}{ratio} &
  \multicolumn{1}{c}{} &
  \multicolumn{1}{c}{seed count} &
  \multicolumn{1}{c}{ratio} &
  \multicolumn{1}{c}{} &
  \multicolumn{1}{c}{seed count} &
  \multicolumn{1}{c}{ratio} &
  \multicolumn{1}{c}{} &
  \multicolumn{1}{c}{seed count} &
  \multicolumn{1}{c}{ratio} \\ \hline
jq               & 1856  & 348  & \textbf{18.75\%} & 1919  & 113  & 5.89\%           & 1850  & 115  & 6.22\%  & 1715  & 137 & 7.99\%          \\
php              & 17199 & 1807 & \textbf{10.51\%} & 16962 & 279  & 1.64\%           & 16485 & 575  & 3.49\%  & 17456 & 436 & 2.50\%          \\
xml              & 6849  & 972  & \textbf{14.19\%} & 6297  & 166  & 2.64\%           & 6437  & 154  & 2.39\%  & 6561  & 149 & 2.27\%          \\
jsoncpp\_fuzzer          & 1321  & 5    & 0.38\%           & 1231  & 1    & 0.08\%           & 1198  & 3    & 0.25\%  & 1283  & 11  & \textbf{0.86\%} \\ \hline
mujs             & 11060 & 3546 & \textbf{32.06\%} & 9093  & 544  & 5.98\%           & 8756  & 603  & 6.89\%  & 9358  & 400 & 4.27\%          \\
ossfuzz          & 8298  & 2995 & \textbf{36.09\%} & 7431  & 1274 & 17.14\%          & 7183  & 1275 & 17.75\% & 7241  & 883 & 12.19\%         \\
cflow            & 1357  & 31   & \textbf{2.28\%}  & 1305  & 8    & 0.61\%           & 1302  & 3    & 0.23\%  & 1253  & 0   & 0.00\%          \\
lua              & 2945  & 1013 & 34.40\%          & 3301  & 1319 & \textbf{39.96\%} & 3015  & 853  & 28.29\% & 2834  & 732 & 25.83\%         \\ \hline
curl             & 2235  & 0    & 0.00\%           & 2177  & 6    & \textbf{0.28\%}  & 2085  & 0    & 0.00\%  & 2178  & 5   & 0.23\%          \\
openssl\_x509    & 2890  & 0    & 0.00\%           & 2915  & 0    & 0.00\%           & 2982  & 0    & 0.00\%  & 2918  & 0   & 0.00\%          \\
base64           & 133   & 3    & \textbf{2.26\%}  & 124   & 1    & 0.81\%           & 128   & 0    & 0.00\%  & 138   & 2   & 1.45\%          \\
md5sum           & 395   & 9    & \textbf{2.28\%}  & 289   & 3    & 1.04\%           & 413   & 2    & 0.48\%  & 262   & 3   & 1.15\%          \\ \hline
\textbf{Average} &       &      & \textbf{12.77\%} &       &      & 6.34\%           &       &      & 5.50\%  &       &     & 4.89\%          \\ \hline
\end{tabular}%
}
\end{table*}

A seed generated by a chat mutator is imported when it triggers new edge coverage that is not yet found by the greybox counterpart of \sys. Note that more interesting seeds could stem from one imported seed by generating new mutations for it. Here we use the imported seed ratio as the metric for assessing the effectiveness and contribution of our AI-augmented mutator.

On average, 12.77\% of the fuzzer queue seeds are imported from the chat mutator in \sys. In particular, the chat mutator contributes to over 30\% of the fuzzer queue for three programs: \texttt{mujs, ossfuzz} and \texttt{lua}. It is worth mentioning that these three programs take source code in various programming languages. \sys contributed over 10\% of the fuzzer queue for another three tested programs: \texttt{jq, php, xml}, which all take formatted data files as input. For the programs in the third category listed in \autoref{tab:benchs}, \sys makes a low to none contribution to the greybox fuzzer. This is because the input format is relatively non-trivial for these programs. Specifically, \texttt{md5sum} takes as input a valid MD5 checksum which is a 32-character hexadecimal number computed on a file. The value appears to be a random sequence of characters compared to XML object which is more close to natural language. It is hard for a natural language-based model like ChatGPT to generate such a value.

\begin{figure*}[th]
    \centering
    \includegraphics[width=\linewidth]{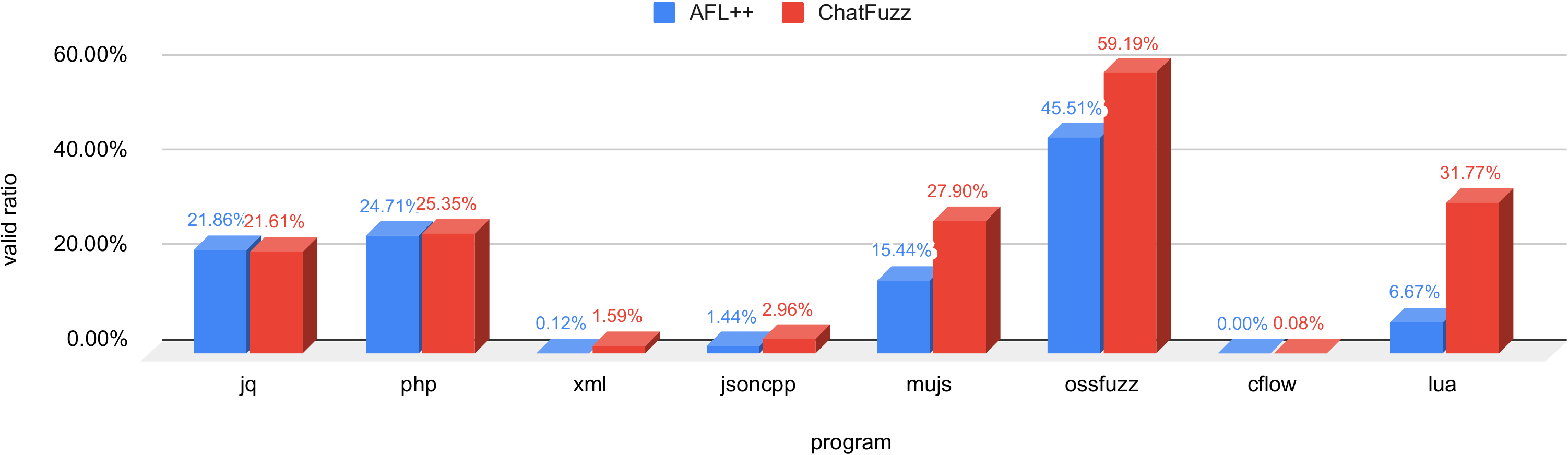}
    
    \caption{Ratio of syntactically valid seeds in fuzzer queue}
    \label{fig:vratio} 
    
\end{figure*}

In ~\autoref{fig:vratio}, we show the ratio of valid seeds in the fuzzer queue of AFL++ and \sys for the dataset, excluding four programs (\texttt{curl, openssl\_x509, base64, md5sum}) of which the input is not of trivial format nor conforms to explicit syntax rules. According to the result, the ratio of syntactically valid seeds in fuzzer queue is higher in \sys than in AFL++ in for all target programs except for \texttt{jq}. For \texttt{jq}, \sys and AFL++ have 393 and 392 valid seeds in the fuzzer queue. For \sys, 28.24\% of the valid seeds are imported from chat mutator. 

The experiment result shows that the chat mutator is able to generate more valid seeds and contribute to the greybox fuzzer.

\subsection{RQ3: Security Application}

In this section, we evaluate the bug detection ability of AFL++ and \sys. Specifically, we inspect the program crashing inputs for six programs including four from UniBench benchmark and two from LAVA-M dataset. In ~\autoref{tab:bugs}, we present the number of unique bugs detected in 8h by each baseline respectively.

\begin{table}[th]
\centering
\caption{Unique Bugs Detected in 8h}
\label{tab:bugs}
{\small%
\begin{tabular}{cccccc}
\hline
\multirow{2}{*}{\textbf{Program}} & \multirow{2}{*}{\textbf{AFL++}}     & \multicolumn{4}{c}{\textbf{\sys}}                                                            \\ \cline{3-6} 
\multicolumn{1}{l}{} & \multicolumn{1}{l}{} & \multicolumn{1}{l}{} & \multicolumn{1}{l}{{\bf -F}} & \multicolumn{1}{l}{{\bf -C}} & \multicolumn{1}{l}{{\bf -CF}} \\ \hline

jq      & - & - & - & - & - \\
mujs    & - & - & - & - & - \\
ossfuzz & - & - & - & - & - \\
cflow   & 2 & \textbf{3} & \textbf{3} & 2 & 1 \\ \hline
base64  & 5  & -  & -  & -  & \textbf{12}  \\
md5sum  & -  & -  & -  & -  & -  \\ \hline
\end{tabular}%
}
\vspace{-0.15in}
\end{table}

According to the result, AFL++ and \sys are not able to trigger any bugs in \texttt{jq, mujs, ossfuzz} or \texttt{md5sum}. For program \texttt{cflow}, \sys is able to find more bugs than AFL++. For program \texttt{md5sum}, \sys-CF is able to find 12 bugs. 

%% file: 5_discuss.tex
\section{Limitations and Discussions}

\medskip
\noindent\textit{\textbf{Unsupported Input Format. }} For now \sys can be utilized for fuzzing programs that execute on text-based formatted input. Other data formats such as image, video, audio or data format that requires particular data loader such as pdf, WOFF or compressed file are not supported. \sys is limited by the AI model used in the chat mutator. With a generative AI model that supports more data format, for instance DALL$\cdot$E, \sys will able to generate testcases in the format of image. For text-based input that does not have explicit syntax rules like hash values, it is hard for generative AI to figure out the underlying rule for the text and therefore the chat mutator could make low to none contribution to the greybox fuzzer.


\medskip
\noindent\textit{\textbf{LLM Model Limitations. }} \sys currently is built with the OpenAI API service which is of high latency compared to a native LLM model. There are other limitations that come along with the API service such as context length, query rate also play a factor in the performance of \sys. In case where the program states can only be triggered by extremely long input, the model is not able to generate that. Moreover, the model does not memorize the past prompts automatically. Therefore, it cannot be directly applied for fuzzing a stateful targets such as network protocol.

\medskip
\noindent\textit{\textbf{Automated Parameter Finetuning. }} Precedent work~\cite{lyu2019mopt} suggests that different mutators should be prioritized dynamically along the fuzzing course to improve the coverage efficiency. The chat mutators under different parameter configurations are practically different mutators. Specifically, when prompting the AI model with low sampling temperature, the difference between the sample input and the new seed is smaller than with a sampling high temperature. Potentially, we can dynamically adjust the parameter configuration of the chat mutator to improve the coverage efficiency. We leave this for future work.




%% file: 6_related.tex
\section{Related Work}
Fuzz testing has gained popularity in both academia and industry due to its black/grey box approach with a low barrier to entry~\cite{fse2014orso}. The key idea of fuzz testing originates from random test generation, where inputs are incrementally produced in the hope of exercising previously undiscovered program behavior~\cite{randoop,csallner2004jcrasher,evosuite}.

\smallskip
\vspace{0.5pt}\noindent\textit{\textbf{Grey-box Fuzzing.}}  
Grey-box fuzzing begins with a seed input, executing the program and iteratively generating new inputs by mutating the previous ones. New inputs are added to the queue if they improve a specified guidance metric such as branch coverage. American Fuzzy Lop is one of the most widely used fuzzing tools~\cite{afl}. Traditional coverage-guided fuzz testing faces challenges in efficiency and effectiveness due to a vast space of inputs and unbounded program paths. Lemieux et al.~tackle this by identifying rarely executed branches with AFL-generated inputs and devising custom mutations to prioritize the exploration of these branches~\cite{fairFuzz}. As a result, it requires fewer fuzzing loops and achieves higher coverage in less time. Other approaches incorporate symbolic execution in fuzzing to guide careful selection and mutation of the inputs, invoking unique program paths~\cite{driller, ChaSp15}. Padhye et al.~introduce Zest~\cite{padhye2019semantic}, which incorporates the semantic validity of input mutations by mapping bit-level changes to valid structural modifications, reducing the search space of inputs.

The effectiveness of fuzz testing heavily relies on the quality of initial seeds. Yet, commonly used seeds tend to traverse similar "high-frequency" paths. To expand path coverage without significantly increasing the number of tests, researchers have designed strategies for seed selections. AFLFast~\cite{bohme2016coverage} models coverage-based greybox fuzzing as a Markov chain, and assigns different selection probabilities for different seeds. EcoFuzz~\cite{yue2020ecofuzz} improves AFLFast's Markov chain model and presents a variant of the Adversarial Multi-Armed Bandit model. EcoFuzz sets three states of the seeds set and develops a unique adaptive scheduling algorithm. 

Most fuzz and random testing techniques are built on the assumption that {\em making small bit/byte level changes to inputs can result in meaningful data}. However, none of these techniques effectively leverages AI or large language models in tandem with
software monitors to guide test input generation, as demonstrated by \sys.

\medskip\noindent\textit{\textbf{Grammar-Based Test Generation.}} 
One challenge in grey-box fuzz testing is generating valid inputs, especially for highly-structured inputs and object-oriented programs. This has led to research efforts to minimize unfruitful fuzzing iterations by generating legal inputs for the target program using input grammars. For example, CodeAlchemist~\cite{han2019codealchemist} is a code generation engine that can systematically generate both syntactically and semantically correct JavaScript code snippets. Token-Level AFL~\cite{Salls2021Token_Level} mutates JavaScript programs at a token level. Tokens from a dictionary are used when mutating programs to replace, insert, or overwrite existing tokens. Le et al.~propose a grammar-based fuzzing approach called Saffron that relies on a user-defined grammar~\cite{Saffron}. During fuzzing, if an input generated by the grammar leads to a program failure, Saffron reconstructs the grammar according to newly learned input specifications of the program. Wang et al.~leverage a user-provided grammar, but instead of arbitrary mutations, they introduce grammar-specific mutations to diversify test inputs for tightly formatted input domains such as XML and JSON~\cite{superion}. Gopinath et al.~highlight that the state-of-art grammar-aware fuzzer {\it dharma}~\cite{mozillasecurity_2020} is still two orders of magnitude slower than a random fuzzer and suggest guidelines for efficient grammar-aware fuzzing~\cite{gopinath2019building}. In their follow-up work, they present an approach to infer an input grammar from the interactions between an input parser and input data~\cite{gopinath2019inferring}.

All existing grammar-based techniques require developers and users to understand the input grammar or manually create data generators. Additionally, fuzzing techniques developed for one type of grammar may not readily translate to other grammar types. In this work, we leverage large language models to automatically infer grammar from seed inputs and generate new valid test cases, mitigating the need for manual grammar specification.

\medskip\noindent\textit{\textbf{AI-Based Testing.}} Previous research has explored the application of AI techniques in program and test case generation~\cite{she2019neuzz,mtfuzz,prefuzz}. For example, NEUZZ~\cite{she2019neuzz} introduces a gradient-descent based approach to fuzz testing by creating a smooth surrogate function to approximate the target program's discrete branching behavior. AthenaTest~\cite{tufano2021unit} trains local transformer-based networks to generate test inputs from a corpus of focal methods and test inputs. Jigsaw~\cite{jigsawICSE2022} is a program synthesis tool based on LLMs and validates their correctness using existing test cases, which diverges from the focus of this paper.

The use of the prompting LLMs has also gained prominence in recent research efforts. Bareiß~et al. employ Codex by providing prompts containing one method-test pair and the method to be tested, enabling the generation of test oracles and test cases~\cite{barei2022code}. ChatUniTest~\cite{xie2023chatunitest} uses ChatGPT to generate Junit test cases. FuzzGPT~\cite{deng2023large} leverages LLMs to generate unusual programs seeking to trigger abnormal behavior in deep learning library APIs. CodaMosa~\cite{lemieux2023codamosa} uses Codex as a black-box tool to generate tests without requiring explicit training, and incorporate these tests into a search algorithm. In contrast, \sys leverages LLMs as mutation operators in the fuzzing process. 


%% file: 7_conclusion.tex
\section{Conclusion}
In this paper, we target the problem of greybox fuzzing for real-world programs that take highly structured data as input. We identify the format-conforming inputs generation to be one major limitation of both mutational-based approach and generative approach. And we propose to utilize the power of generative large language model for generating well-formatted program inputs to improve greybox fuzzing. The idea is implemented as a prototype \sys and evaluated on 12 target programs across three benchmarks extensively tested in precedent works. Result proves that our approach has higher coverage efficiency than SOTA greybox fuzzer AFL++. And it is able to generate more syntactically valid seeds that explore deeper program space.